\documentclass[conference]{IEEEtran}
\IEEEoverridecommandlockouts
\usepackage{multicol}
\usepackage{comment}
\usepackage{cite}
\usepackage{hyperref}
\usepackage{lipsum}
\usepackage{adjustbox}
\usepackage{amsmath,amssymb,amsfonts}
\usepackage{float}
\usepackage{algorithmic}
\usepackage{pgfplots}
\usepackage{graphicx}
\usepackage{textcomp}
\usepackage{xcolor}
\usepackage{hyperref}
\usepackage{comment}

\def\BibTeX{{\rm B\kern-.05em{\sc i\kern-.025em b}\kern-.08em
    T\kern-.1667em\lower.7ex\hbox{E}\kern-.125emX}}

\newcommand{\squishlist}{
   \begin{list}{$\bullet$}
    { \setlength{\itemsep}{0pt}      \setlength{\parsep}{0pt}
      \setlength{\topsep}{-3pt}       \setlength{\partopsep}{0pt}
      \setlength{\listparindent}{-2pt}
      \setlength{\itemindent}{-5pt}
      \setlength{\leftmargin}{1em} \setlength{\labelwidth}{0em}
      \setlength{\labelsep}{0.5em} } }

\newcommand{\squishend}{
    \end{list}  }

\begin{document}

\title{Efficient Implementation of Multi-Channel Convolution in Monolithic 3D ReRAM Crossbar}

\author{\IEEEauthorblockN{Sho Ko}
\IEEEauthorblockA{\textit{School of ECE} \\
\textit{Georgia Tech}\\
Atlanta, GA, USA \\
sko.45@gatech.edu}
\and
\IEEEauthorblockN{Yun Joon Soh}
\IEEEauthorblockA{\textit{Department of CSE} \\
\textit{UC San Diego}\\
La Jolla, CA, USA \\
yjsoh@eng.ucsd.edu}
\and
\IEEEauthorblockN{Jishen Zhao}
\IEEEauthorblockA{\textit{Department of CSE} \\
\textit{UC San Diego}\\
La Jolla, CA, USA \\
jzhao@eng.ucsd.edu}\\
}
\maketitle

\begin{abstract}
Convolutional neural networks (CNNs) demonstrate promising accuracy in a wide range of applications. Among all layers in CNNs, convolution layers are the most computation-intensive and consume the most energy. As the maturity of device and fabrication technology, 3D resistive random access memory (ReRAM) receives substantial attention for accelerating large vector-matrix multiplication and convolution due to its high parallelism and energy efficiency benefits. However, implementing multi-channel convolution naively in 3D ReRAM will either produce incorrect results or exploit only partial parallelism of 3D ReRAM. In this paper, we propose a 3D ReRAM-based convolution accelerator architecture, which efficiently maps multi-channel convolution to monolithic 3D ReRAM. Our design has two key principles. First, we exploit the intertwined structure of 3D ReRAM to implement multi-channel convolution by using a state-of-the-art convolution algorithm. Second, we propose a new approach to efficiently implement negative weights by separating them from non-negative weights using configurable interconnects. Our evaluation demonstrates that our mapping scheme in 16-layer 3D ReRAM achieves a speedup of 5.79$\times$, 927.81$\times$, and 36.8$\times$ compared with a custom 2D ReRAM baseline and state-of-the-art CPU and GPU. Our design also reduces energy consumption by 2.12$\times$, 1802.64$\times$, and 114.1$\times$ compared with the same baseline.
\end{abstract}


\section{Introduction}
Deep learning algorithms are adopted in a wide range of systems, whether small edge devices or large data centers~\cite{b1}. Convolutional neural networks (CNNs) have revolutionized deep learning applications by achieving unprecedented accuracy for object detection and image classification. However, CNNs are time-consuming and power-hungry during the computation process. For example, AlexNet~\cite{b18} performs $10^9$ operations for a single image input without batching~\cite{b5}. Convolution layers are the most computation-demanding in CNNs. It is estimated that the convolution layers of VGG-16~\cite{b16} take $67.8\%$ of the total execution time~\cite{b21}.

\begin{figure}[t]
\centering
\includegraphics[width=60mm]{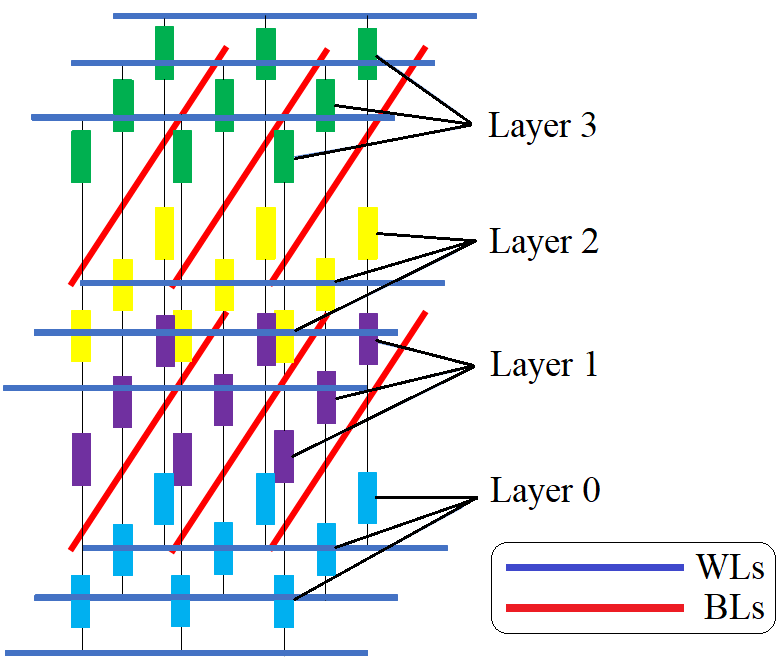}
\vspace{-5pt}
\caption{Horizontally integrated monolithic 3D ReRAM.}
\vspace{-20pt}
\label{fig:3D}
\end{figure}

Recently, resistive random access memory (ReRAM) is becoming an attractive technology solution for accelerating convolution layers, due to its promising parallelism and energy efficiency benefits. ReRAM is a novel memory technology which consists of a crossbar structure of memristors. It combines storage and computation together and accelerates deep neural networks in the analog domain. Recently, several ReRAM-based processing-in-memory (PIM) accelerators have been proposed such as PRIME~\cite{b3}, ISAAC~\cite{b4}, and PipeLayer~\cite{b5}. These architectures all focused on efficiently architecting 2D ReRAM for CNN applications. However, monolithic 3D integration technology has grown rapidly. Compared with 2D ReRAM, 3D ReRAM can provide more parallelism, take less area, produce less noise, and consume less energy in computations~\cite{b8}. Monolithic 3D ReRAM can be either vertically integrated or horizontally integrated. In our work, we focus on mapping multi-channel convolution to horizontally integrated monolithic 3D ReRAM because it can be more reliably fabricated~\cite{b8}, as shown in Fig.~\ref{fig:3D}.

Nonetheless, leveraging 3D ReRAM for processing multi-channel convolution in parallel still faces three challenges. First, even though previous works have successfully designed several accelerators which used 2D ReRAM to process multi-channel convolution, simply extending 2D ReRAM to 3D ReRAM without any modification will produce incorrect results due to the stacked structure. Second, even multi-channel convolution can be correctly implemented, a naive implementation will exploit only partial parallelism of 3D ReRAM. Third, kernels in multi-channel convolution, like edge detection filters, sometimes has negative weights. An efficient way to implement negative weights in 3D ReRAM is necessary.

Our goal in this paper is to efficiently map multi-channel convolution to horizontally integrated monolithic 3D ReRAM. In order to achieve our goal, we propose an convolution accelerator with two design principles. First, to solve challenges 1 and 2, we for the first time exploit the massive parallelism of 3D ReRAM to accelerate CNNs by using a newly proposed algorithm to implement multi-channel convolution~\cite{b10}. Second, to solve challenge 3, we propose a new approach to efficiently implement negative weights in 3D ReRAM using configurable interconnects.

\section{Background}
In this section, we describe ReRAM background and motivate our design.

\subsection{Convolutional Neural Networks}
\begin{figure}[h!]
\centering
\includegraphics[width=80mm]{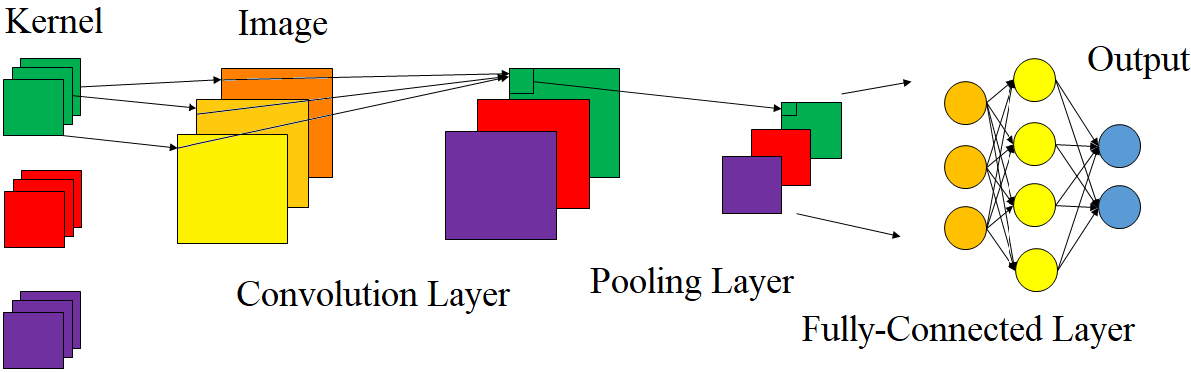}
\vspace{-15pt}
\caption{Convolutional neural network.}
\vspace{-10pt}
\label{fig:cnn}
\end{figure}

CNNs are the heart of current deep learning applications. A typical CNN consists of multiple layers, such as convolution layers, pooling layers, and fully-connected layers, as shown in Fig.~\ref{fig:cnn}.

\vspace{-10pt}

\subsection{Memristor and 2D ReRAM}
\begin{figure}[b]
\vspace{-20pt}
\centering
\includegraphics[width=50mm]{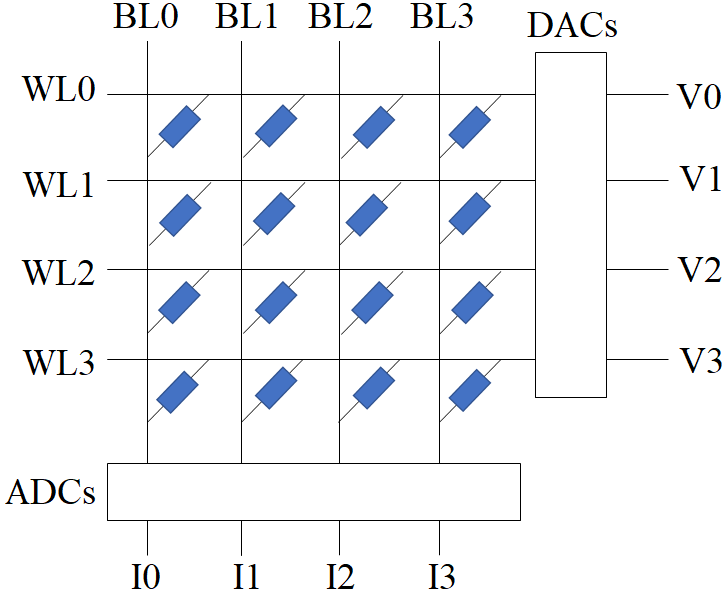}
\vspace{-10pt}
\caption{2D ReRAM crossbar for vector-matrix multiplication.}
\label{fig:2D}
\end{figure}

2D ReRAM is a grid structure consisting of multiple memristors, as shown in Fig.~\ref{fig:2D}. Each ReRAM cell contains one memristor. Such design can exploit the analog characteristics of ReRAM to perform fast and energy-efficient matrix multiplication and convolution. Vector-matrix multiplication can be easily calculated using ReRAM, because of two basic electrical theorems, Ohm's law and Kirchhoff's current law. Ohm's law states that the current through a resistor is equal to the voltage across the resistor divided by the resistance of the resistor ($I=V/R$), which is also equal to the voltage across the resistor multiplied by the conductance of the resistor ($I=VG$). This law makes performing analog floating-point multiplication possible. Kirchhoff's current law states that the total current output is equal to the sum of all input current for a node in the circuit. This law makes performing analog floating-point addition possible. Vector-matrix multiplication can be mapped to ReRAM in the following three steps, as shown in Fig.~\ref{fig:2D}: First, the digital input is converted to analog signals by digital-to-analog converters (DACs) and then mapped to the voltage on horizontal bit lines (WLs); Second, the weight matrix is quantized and then mapped to the conductance of memristors; Third, the output signals are read from the current on the vertical bit lines (BLs) and then converted to digital output by analog-to-digital converters (ADCs).

\subsection{Monolithic 3D ReRAM}
Monolithic 3D ReRAM integrates ReRAM cells in either a vertical or a horizontal manner. For example, B. Chakrabarti \textit{et al.} developed an 8-layer vertically integrated monolithic 3D ReRAM~\cite{b7}, while M. Mao \textit{et al.} designed a horizontally integrated monolithic 3D ReRAM~\cite{b8}. Horizontally integrated monolithic 3D ReRAM has more reliable manufacturing technology~\cite{b8}. An example of horizontal 3D ReRAM is shown in Fig.~\ref{fig:3D}. Its intertwined structure ensures that WLs and BLs between adjacent layers are shared. The 4-layer 3D ReRAM has three voltage planes with three WLs on each plane. It also has two current planes with three BLs on each plane. Different layers of memristors have different colors.

Such 3D structure has several advantages compared with the 2D structure. First, 3D ReRAM has less area than the 2D version for the same amount of memristors. Second, 3D ReRAM has shorter WLs and BLs to avoid parasitic resistance which may introduce unnecessary noise in the circuit and compromise the output integrity~\cite{b6}. Third, shared WLs and BLs between adjacent layers in 3D ReRAM lead to better utilization of peripheral circuits to save space and energy~\cite{b9}. Finally, shared BLs connect two adjacent layers of memristors. According to Kirchhoff's current law, the current on the BL is equal to the sum of the current from the two adjacent layers. It is represented as
\begin{equation}
I=V_{above}G_{above}+V_{below}G_{below}
\end{equation}
in Fig.~\ref{fig:3D}. This property is helpful for mapping convolution to 3D ReRAM and maximize computational parallelism. Motivated by Y. Huang \textit{et al.}, who leveraged the massive parallelism of monolithic 3D ReRAM for graph processing algorithms~\cite{b9}, we design a convolution accelerator which uses the same structure for multi-channel convolution.

\section{Proposed Approach}

In order to exploit the massive parallelism provided by monolithic 3D ReRAM to implement the computation-intensive multi-channel convolution layers in CNNs, we propose a convolution accelerator, which efficiently maps multi-channel convolution to monolithic 3D ReRAM.

\subsection{Architecture Design}
\begin{figure}[t]
\centering
\includegraphics[width=85mm]{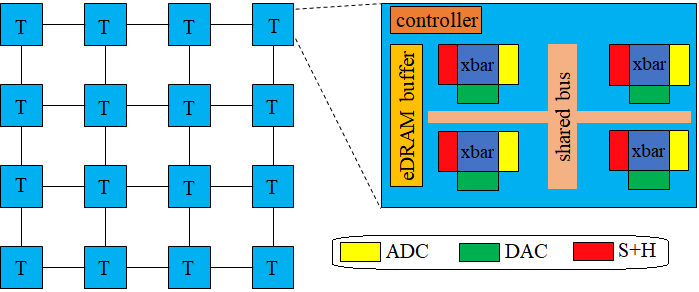}
\vspace{-10pt}
\caption{Overall architecture.}
\vspace{-15pt}
\label{fig:arch}
\end{figure}

We employ an optimized architecture for accelerating CNNs, as shown in Fig.~\ref{fig:arch}. The architecture is composed of multiple tiles of ReRAM cells connected by an on-chip mesh~\cite{b4}. Each tile has a eDRAM buffer, a shared bus, a controller, and multiple ReRAM-based processing engines. We substitute the conventional 2D ReRAM crossbar with monolithic 3D ReRAM crossbar. Each processing engine communicates with the buffer via the shared bus. The controller maps multi-channel convolution to the processing engines and helps configure interconnects.

\subsection{Convolution Algorithm}
Common 2D convolution algorithms include single kernel single channel (SKSC), single kernel multiple channel (SKMC), and multiple kernel multiple channel (MKMC). MKMC is widely used in many CNN architectures. In order to present how MKMC without batching works, we define image to be $I$ and kernel to be $K$. $I$ is a 3D matrix with dimensions $c (channel) \times h (height) \times w (width)$. $K$ is a 4D matrix with dimensions $n (kernel) \times c (channel) \times l (length) \times l (length)$. SKSC is a simple convolution between one channel of the image and one channel of one kernel, which is defined as
\begin{equation}
SKSC(I_{i},K_{j,i})=conv(I_{i},K_{j,i})
\end{equation}
where $i \in [0,c)$ and $j \in [0,n)$. SKMC is calculated by summing the result of SKSC of every corresponding channel of the image and one kernel, which is defined as
\begin{equation}
SKMC(I,K_{j})=\sum_{i=0}^{c-1} conv(I_{i},K_{j,i})
\end{equation}
where $j \in [0,n)$. MKMC is computed by concatenating the result of SKMC of the image and every kernel, which is defined as
\begin{equation}
MKMC(I,K)=SKMC(I,K_{0})|\cdots|SKMC(I,K_{n-1})
\end{equation}
where $|$ represents concatenation.

\begin{figure}[t]
\centering
\includegraphics[width=85mm]{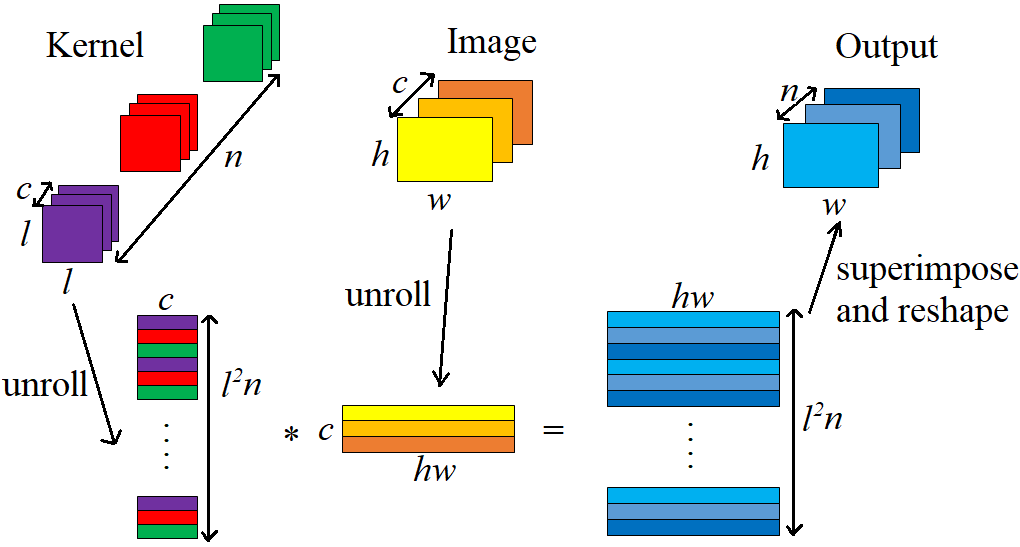}
\vspace{-10pt}
\caption{Computation process of MKMC~\cite{b10}.}
\vspace{-20pt}
\label{fig:MKMC}
\end{figure}

Traditionally, MKMC is calculated by unrolling each kernel into a row vector in the kernel matrix and corresponding image pixels to a column vector in the image matrix. Then multiplying the two matrices gives the result. However, this approach is not suitable for mapping to 3D ReRAM because the property in equation (1) cannot be easily applied. Recently, new approaches to compute MKMC have been proposed~\cite{b10}. One of them is to compute the $n^2$ convolution using $n^2$ different $1\times1$ convolutions. It unrolls the corresponding $1\times1$ weights in all channels within one kernel into a row vector in the kernel matrix and the corresponding $1\times1$ pixels in all channels within the image into a column vector. After multiplying the two matrices, there are $l^2$ submatrices of size $n\times(h\times w)$. We need to superimpose the submatrices together into one matrix and reshape it to be $h\times w \times n$, as shown in Fig.~\ref{fig:MKMC}. This algorithm is suitable for mapping to 3D ReRAM. In particular, the superimposition step can be efficiently implemented using the property in equation (1).

\subsection{Efficient Mapping of Convolution to 3D ReRAM}
We design an efficient mapping of the algorithm to 3D ReRAM. We start with mapping the $n \times c \times l \times l$ kernel to 3D ReRAM. We employ $c$ WLs in each voltage plane and $n$ BLs in each voltage plane. Since we use 3D ReRAM with shared WLs and BLs, the number of layers has to be an even number for reconfiguration. If $l^2$ is an even number, we use $l^2$ layers of memristors, $\frac{l^2}{2}+1$ voltage planes, and $\frac{l^2}{2}$ current planes. If $l^2$ is an odd number, we use $l^2+1$ layers of memristors, $\frac{l^2+1}{2} + 1$ voltage planes, and $\frac{l^2+1}{2}$ current planes. Note when $l^2$ is odd, one layer of memristors is not in use (dummy layer). We need to either set the resistance of the memristors close to zero or set the voltage on the relevant WL to zero to ensure correct output current. For each voltage plane, $c$ WLs correspond to one column of the image matrix. WLs from different voltage planes but on the same vertical plane have the same voltage to maximize parallelism of the 3D structure. One column of the image matrix can be fed into 3D ReRAM at one logical cycle. It takes $h \times w$ logical cycles to pass the $c \times h \times w$ image into 3D ReRAM. For each current plane, $n$ BLs correspond to $n$ kernels in the output. BLs from different current planes but on the same vertical plane are accumulated simultaneously to implement the superimposition. 3D ReRAM produces $n$ sums at one logical cycle. After $h \times w$ logical cycles, it will produce the $n \times h \times w$ output.

\begin{figure}[b]
\vspace{-20pt}
\centering
\includegraphics[width=85mm]{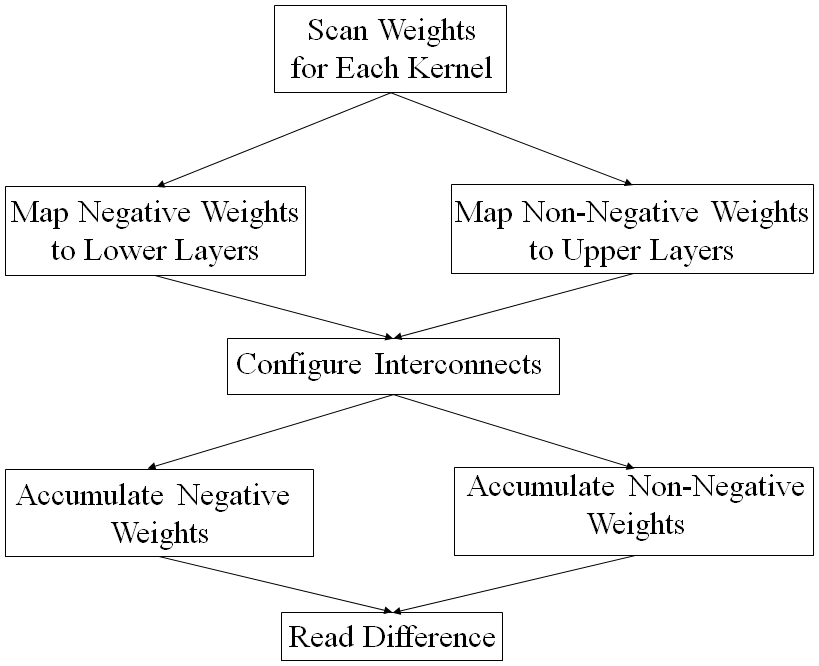}
\vspace{-10pt}
\caption{Flow diagram of the full mapping scheme.}
\label{fig:diagram}
\end{figure}

In addition, we present a new approach to implement negative weights using configurable interconnects. The full mapping scheme is summarized in the flow diagram in Fig.~\ref{fig:diagram}. First, we scan each of $n$ kernels and count the number of negative weights and non-negative weights. Note there is a voltage plane that separates negative weights from non-negative weights for each kernel. Second, negative weights are mapped to the lower layers below the voltage separation plane and non-negative weights are mapped to the upper layers above the voltage separation plane. Third, interconnects are configured for negative weights and non-negative weights accordingly. Fourth, negative weights and non-negative weights are accumulated separately and then fed into peripheral circuits which are used to read the difference between the two accumulated currents.

\subsection{Putting It All Together: An Example}
We demonstrate our approach using an example of applying an edge detection filter to an image with three channels. The filter has two kernels each with three channels of the same value, as shown in Fig. 7(a)-(b). We use a 10-layer 3D ReRAM (0 to 9) with six voltage planes (0 to 5) and five current planes (0 to 4). For kernel 0, we set the voltage on voltage plane 5 to zero because we do not use memristors in layer 9. We use four layers (0 to 3) for negative weights and five layers (4 to 8) for non-negative weights. The separation plane is voltage plane 2. After configuring interconnects, we accumulate two current planes (0 to 1) for negative weights as $I_{n}$ and three current planes (2 to 4) for non-negative weights as $I_{p}$, as shown in Fig. 7(c). For kernel 1, we set the voltage on voltage plane 0 to zero because we do not use memristors in layer 0. We use one layer (1) for negative weights and eight layers (2 to 9) for non-negative layers. The separation plane is voltage plane 1. After configuring interconnects, we accumulate one current plane (0) for negative weights as $I_{n}$ and four current planes (1 to 4) for non-negative weights as $I_{p}$, as shown in Fig. 7(d). In order to read the current difference, we slightly modify the typical inverting operational amplifier circuit, as shown in Fig. 7(e). Measuring the output current $I_{2}$ gives the difference between $I_{p}$ and $I_{n}$.

We prove the correctness of the circuit in Fig. 7(e). Since the current into the negative input of the op amp is zero, Kirchhoff's current law gives $I_{0}=I_{n}$. Then Ohm's law states $V_{0}=I_{n}R_{0}$. Using Kirchhoff's voltage law, $V_{1}=-I_{n}R_{0}$ holds, and then $I_{1}=-I_{n}$ is true according to Ohm's law again. Finally, we reach the conclusion that $I_{2}=I_{p}-I_{n}$ after applying Kirchhoff's current law again.

\section{Evaluation}
In this section, we evaluate the performance and energy consumption of our proposed design.

\begin{figure}[t]
\centering
\includegraphics[width=85mm]{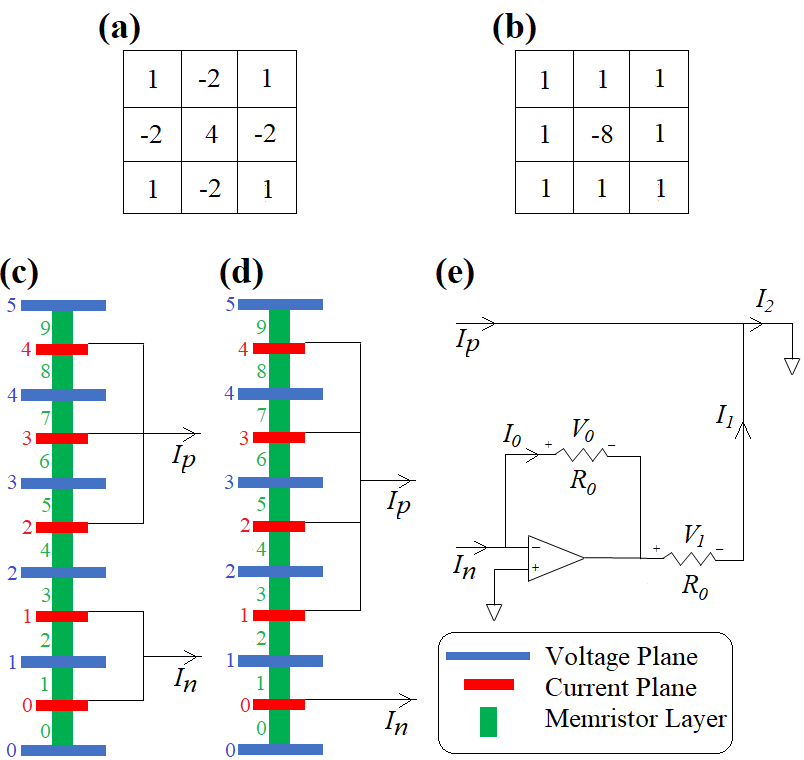}
\vspace{-5pt}
\begin{flushleft}
\footnotesize{Fig. 7. (a) One channel of kernel 0. (b) One channel of kernel 1. (c) Interconnects configuration for kernel 0. (d) Interconnects configuration for kernel 1. (e) Circuit diagram of inverting operational amplifier.}
\end{flushleft}
\vspace{-20pt}
\end{figure}

We first compare the performance of 2D ReRAM with other popular memory technologies such as SRAM, eDRAM, PCM, and STT-RAM. We use DESTINY tool~\cite{b12} to simulate 1 GB of each type of memory using 32 nm technology. Table I shows the parameters of each memory type. It is easy to observe that ReRAM performs better in read/write energy and read/write latency than eDRAM and SRAM. In addition, compared with STT-RAM, ReRAM has smaller read/write energy and read latency at the expense of larger write latency.

We then explore monolithic 3D ReRAM by evaluating the relationship between number of layers in 3D ReRAM and its performance. Again, we use DESTINY tool~\cite{b12} to simulate 1 GB of 3D ReRAM using 32 nm technology. We use the read/write latency and energy of 2-layer 3D ReRAM as baseline and normalize 3D ReRAM with more layers to it, as shown in Fig. 8. We observe that for the same memory capacity, as the number of layers increases, read/write energy and read/write latency also increase. In our experiment, we use profiling results to optimize the number of layers in 3D ReRAM to balance between more parallelism versus higher read/write latency and energy.

\vspace{-5pt}
\begin{table}[h!]
\small{TABLE I: Parameters of several memory types.}
\vspace{-5pt}
\centering
\begin{tabular}[t]{|l|c|c|c|c|}
\hline
 & Write & Read & Write & Read \\
 & Energy & Energy & Latency & Latency \\
 & ($nJ$) & ($nJ$) & ($ns$) & ($ns$) \\
\hline
ReRAM & 1.907 &	1.623 &	15.274 & 13.948 \\
eDRAM & 3.407 &	3.324 &	34.207 & 66.661 \\
SRAM & 6.687 &	6.688 &	144.556 & 279.546 \\
STT-RAM & 2.102 & 1.975 & 13.469 & 18.06 \\
\hline
\end{tabular}
\end{table}
\vspace{-10pt}

\subsection{Experiment Setup}

\noindent\textbf{Configuration and Simulation.} In our experiment, we use 3D ReRAM with 16 layers for two reasons. First, 16 layers are enough to handle a typical kernel size 3$\times$3. Second, it provides the optimal latency based on the extended report of DESTINY~\cite{b14}. We sacrifice the parallelism to support larger kernels, such as 5$\times$5. If we had smaller number of layers such as 10 or 12, we must repeat the computation more than twice to support the larger kernel. For ReRAM crossbars, we use DESTINY~\cite{b12} to measure the execution time and energy consumption. For interconnects, we model with CACTI 6.5~\cite{b13} at 32 nm. For DACs and ADCs, we use the results from B. Murmann, ``ADC Performance Survey''~\cite{b15}.

\noindent\textbf{Workload and Baseline.} We benchmark several selected MKMC layers from the inference phase of three popular CNN architectures, VGG-16~\cite{b16}, GoogLeNet~\cite{b17}, and AlexNet~\cite{b18}. All three architectures are trained in Tensorflow framework~\cite{b20} and evaluated on the widely used ImageNet database~\cite{b23}. For our baseline, we do not use experimental results from previous works for two reasons. First, it is unfair to compare the performance with different deign focuses. Second, it is difficult to obtain all the detailed design parameters from previous papers.

Instead, we compare the execution time and energy consumption of our design with a custom 2D ReRAM baseline, a CPU platform, and a GPU platform. We implement MKMC using this algorithm~\cite{b10} for the CPU and GPU platform. For the custom 2D ReRAM baseline, we assume 2D ReRAM crossbars in the same architecture with same amount of memristors as our proposed 3D ReRAM design for fair comparison. For the CPU platform, we choose Intel Core i7-5700HQ processor, which has 4 cores, 6 MB cache, and operates around 2.7 GHz. For the GPU platform, we choose NVIDIA GeForce GTX 1080 Ti, which has 3584 CUDA cores, 11 GB GDDR5X graphics memory, and operates around 1582 MHz. The CPU and GPU execution time is measured within the framework. The CPU energy consumption is estimated by Intel Product Specifications~\cite{b19} and the GPU energy consumption is estimated by NVIDIA System Management Interface ({\fontfamily{lmtt}\selectfont nvidia-smi}).

\vspace{5pt}

\begin{center}
\begin{tikzpicture}
\begin{axis}[
    xlabel={\# of layers in 3D ReRAM},
    xmin=1, xmax=33,
    ymin=0.8, ymax=2.7,
    legend pos=north west,
    ymajorgrids=true,
    grid style=dashed,
]
 
\addplot
coordinates {
(2,1)
(4,1.077324478)
(6,1.15512334)
(8,1.268975332)
(10,1.316888046)
(12,1.364326376)
(14,1.412239089)
(16,1.459677419)
(18,1.50711575)
(20,1.555028463)
(22,1.602466793)
(24,1.650379507)
(26,1.697817837)
(28,1.74573055)
(30,1.79316888)
(32,1.840607211)
};

\addplot
coordinates {
(2,1)
(4,1.020251779)
(6,1.041050903)
(8,1.243021346)
(10,1.263820471)
(12,1.284619595)
(14,1.305418719)
(16,1.325670498)
(18,1.346469622)
(20,1.367268747)
(22,1.388067871)
(24,1.408866995)
(26,1.429118774)
(28,1.449917898)
(30,1.470717022)
(32,1.491516147)
};

\addplot
coordinates {
(2,1)
(4,1.116699958)
(6,1.250832274)
(8,1.289268204)
(10,1.354155317)
(12,1.423400521)
(14,1.497003813)
(16,1.574965196)
(18,1.657284668)
(20,1.74396223)
(22,1.835058411)
(24,1.930452152)
(26,2.030203983)
(28,2.134313904)
(30,2.242781914)
(32,2.355608014)
};

\addplot
coordinates {
(2,1)
(4,1.110159193)
(6,1.237817475)
(8,1.456981407)
(10,1.518653542)
(12,1.584639689)
(14,1.655061368)
(16,1.729797059)
(18,1.808968283)
(20,1.892453518)
(22,1.980374286)
(24,2.072609066)
(26,2.169279378)
(28,2.270263702)
(30,2.375622797)
(32,2.485417426)
};

\legend{Write Energy, Read Energy, Write Latency, Read Latency}
\end{axis}
\end{tikzpicture}
\vspace{-10pt}
\begin{flushleft}
\footnotesize{Fig. 8. Normalized read/write latency and energy for monolithic 3D ReRAM with different number of layers.}
\end{flushleft}
\end{center}
\vspace{-10pt}

\subsection{Performance Results}
Fig. 9(a) compares the performance of 3D ReRAM with a custom 2D ReRAM baseline, a CPU platform, and a GPU platform. We use the CPU performance as the baseline and normalize GPU, 2D ReRAM, and 3D ReRAM to it. The speedup of 3D ReRAM compared with 2D ReRAM, CPU, and GPU are 5.79$\times$, 927.81$\times$, and 36.8$\times$, respectively. 3D ReRAM achieves the same inference accuracy as our baseline. Although 3D ReRAM has slightly larger read/write latency than 2D ReRAM, the massive parallelism that 3D ReRAM provides compensates this disadvantage and computes multi-channel convolution faster. In addition, 2D ReRAM doesn't have shared WLs and BLs, resulting in more complex interconnects and longer computation time. 3D ReRAM also achieves significant speedup compared with CPU and GPU because convolution layers require intensive computations and easily tie up digital processors with constant memory access and data movement.

\subsection{Energy Results}
Fig. 9(b) compares the energy consumption of 3D ReRAM with a custom 2D ReRAM baseline, a CPU platform, and a GPU platform. We use the CPU energy consumption as the baseline and normalize GPU, 2D ReRAM, and 3D ReRAM to it. The energy saving of 3D ReRAM compared with 2D ReRAM, CPU, and GPU are 2.12$\times$, 1802.64$\times$, and 114.1$\times$, respectively. 3D ReRAM consumes less energy compared to 2D ReRAM because shared WLs and BLs reduce roughly half digital-to-analog and analog-to-digital computations. It also benefits from less complex interconnects. 3D ReRAM can achieve huge energy reduction compared to CPU and GPU due to two reasons. First, 3D ReRAM uses the analog properties to compute vector-matrix multiplication, which is more energy efficient than most digital computations. Second, 3D ReRAM passes data through stacked layers, which is shorter than the data movement between processing units and memory hierarchy in most digital processors.

\section{Related Works}
Recently, several architectures have been proposed to use ReRAM to accelerate CNN applications. PRIME~\cite{b3}, ISAAC~\cite{b4}, and PipeLayer~\cite{b5} demonstrate the promising performance gain when off-loading the CNN computation to 2D ReRAM crossbar. Our paper contributes in similar but yet different aspects. First, 2D ReRAM has limited parallelism in computation, while our work extends the structure to 3D with shared WLs and BLs to fully exploit its computational capability to process multi-channel convolution layers. However, since our design focus is different from previous works, we cannot use their works as our baseline. Instead, we compare our design with a custom 2D ReRAM baseline and state-of-the-art CPU and GPU. Second, kernel mapping is not efficiently addressed in PRIME, while we present a more efficient mapping based on a recently proposed approach to compute MKMC~\cite{b10}.

\begin{figure*}[t]
\centering
\includegraphics[width=0.9\textwidth,height=10.9cm]{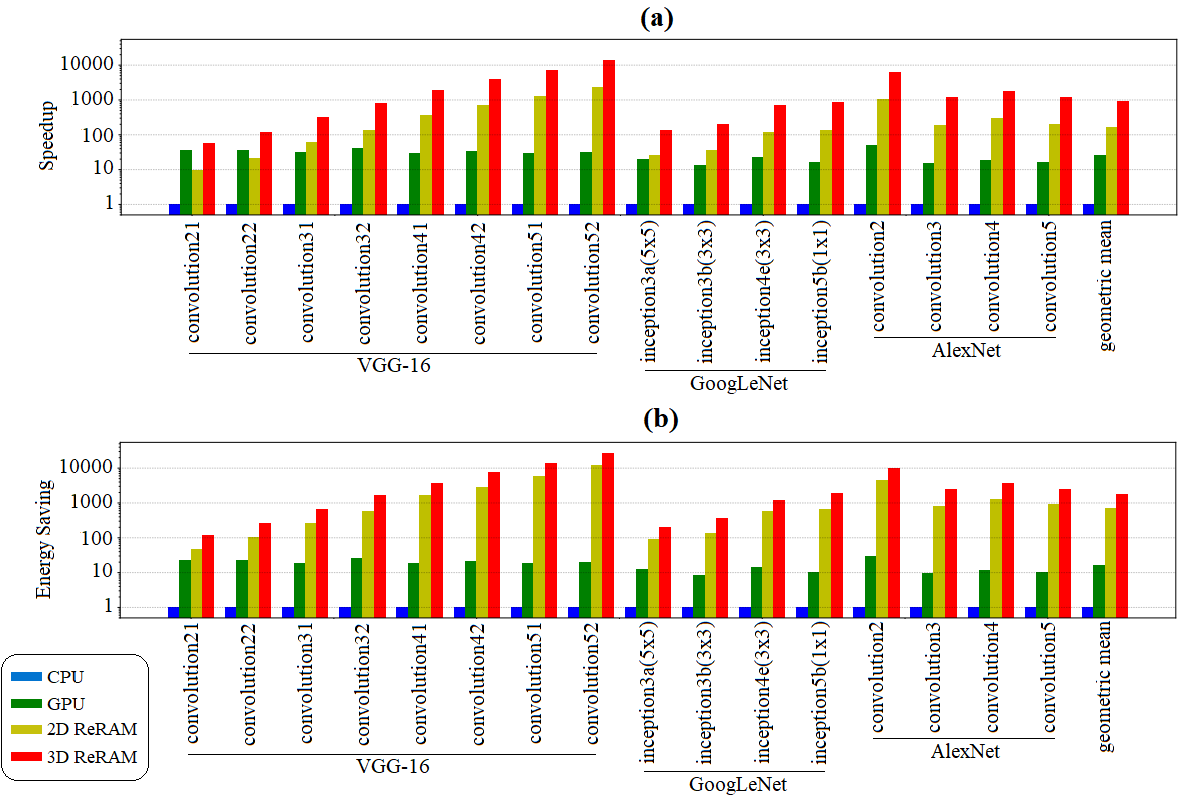}
\begin{flushleft}
\footnotesize{Fig. 9. (a) Normalized 3D ReRAM speedup against 2D ReRAM, CPU, and GPU. (b) Normalized 3D ReRAM energy saving against 2D ReRAM, CPU, and GPU.}
\end{flushleft}
\vspace{-20pt}
\end{figure*}

\section{Conclusion}
This paper presents a convolution accelerator which efficiently maps multiple kernel multiple channel convolution to monolithic 3D ReRAM. By using a newly proposed algorithm, we for the first time take advantage of the property in equation (1) and maximize parallelism of 3D ReRAM to improve the performance and energy efficiency of convolution layers in convolutional neural networks. In order to implement negative weights, we present a new approach to accumulate negative weights and non-negative weights separately using configurable interconnects and calculate the final results using peripheral circuits. Our experiment demonstrates that the proposed mapping achieves a speedup of 5.79$\times$, 927.81$\times$, and 36.8$\times$ compared with a custom 2D ReRAM baseline and state-of-the-art CPU and GPU. Our design also reduces energy consumption by 2.12$\times$, 1802.64$\times$, and 114.1$\times$ compared with the same baseline.

\end{document}